\documentclass{article}
\usepackage{IEEEtrantools}
\usepackage{cite}
\usepackage{amsmath,graphicx}
\usepackage{spconf,amsmath,graphicx}
\usepackage{multirow}
\usepackage{mathtools,xparse}
\usepackage{graphicx}
\usepackage[dvips]{epsfig}
\usepackage{adjustbox}
\usepackage{subcaption}
\usepackage{graphicx}
\usepackage{epsfig}
\usepackage{array}
\usepackage{makecell}
\usepackage{multicol}
\usepackage{multirow}
\usepackage{amssymb,amsmath}
\usepackage{hhline}
\usepackage{verbatim}
\usepackage{amsmath}
\usepackage{mathabx}
\usepackage{caption}
\usepackage{makecell}
\usepackage{url,hyperref}

\usepackage{tabularx}
\usepackage{float}
\floatstyle{plaintop}
\restylefloat{table}
\usepackage[font=small,labelfont=bf]{caption}

\usepackage{caption}
\usepackage[T1]{fontenc}
\usepackage{nicefrac}
\usepackage{color}
\usepackage{subcaption}
\newcommand{\eunwooedit}[1]{\textcolor{black}{#1}}
\newcommand{\ryuichiedit}[1]{\textcolor{black}{#1}}
\newcommand{\minjaeedit}[1]{\textcolor{black}{#1}}
\title{Improving LPCNet-based Text-to-Speech with \\ Linear Prediction-structured Mixture Density Network}
\name{Min-Jae Hwang$^{1, 2}$, Eunwoo Song$^{3}$, Ryuichi Yamamoto$^{4}$, Frank Soong$^{5}$ and Hong-Goo Kang$^{2}$ 	    \thanks{Work partially performed when the first author was an intern at MSRA.}}
\address{$^{1}$Search Solutions Inc., Seongnam, Korea, $^{2}$Yonsei Univ., Seoul, Korea\\
	$^{3}$NAVER Corp., Seongnam, Korea,  $^{4}$LINE Corp., Tokyo, Japan \\
	$^{5}$Microsoft Research Asia, Beijing, China     \vspace{-2mm}}
\begin{document}
	\maketitle
	\fontsize{9.05}{10.5}\selectfont
	\begin{abstract}
	    \vspace{-0mm}
        In this paper, we propose an improved LPCNet vocoder using a linear prediction (LP)-structured mixture density network (MDN).
        The recently proposed LPCNet vocoder has successfully achieved high-quality and lightweight speech synthesis systems by combining a vocal tract LP filter with a WaveRNN-based vocal source (i.e., excitation) generator.
        However, the quality of synthesized speech is often unstable because the vocal source component is insufficiently represented by the $\mu$-law quantization method, and the model is trained without considering the entire speech production mechanism.
        To address this problem, we first introduce LP-MDN, which enables the autoregressive neural vocoder to structurally represent the interactions between the vocal tract and vocal source components.
        Then, we propose to incorporate the LP-MDN to the LPCNet vocoder by replacing the conventional discretized output with continuous density distribution.
        The experimental results verify that the proposed system provides high quality synthetic speech by achieving a mean opinion score of 4.41 within a text-to-speech framework.
	\end{abstract}
	\begin{keywords}
		Speech synthesis, text-to-speech, neural vocoder, LPCNet, LP-MDN
	\end{keywords}
	
	\section{Introduction}
	\label{sec:intro}
	\vspace{-2mm}
	Generative models for raw speech waveform have significantly improved the quality of speech synthesis systems \cite{ze2013statistical, Oord2016WaveNetAG}.
	Specifically, by conditioning acoustic features to the network input, neural vocoding models such as WaveNet, WaveRNN, and WaveGlow have successfully replaced the traditional parametric vocoders \cite{Oord2016WaveNetAG, kalchbrenner2018efficient, prenger2019waveglow}.

	To further improve the performance of speech synthesis system, a recently proposed LPCNet vocoder takes advantages of the merits of both the source-filter model-based parametric vocoder and the WaveRNN structure \cite{valin2019lpcnet, kons2019high}.
	In this framework, a linear prediction (LP) inverse filter is used to decouple the formant-related spectral structure from the input speech signal, and the probability distribution of its residual signal, that is, the excitation signal, is then modeled using the WaveRNN \cite{kalchbrenner2018efficient}. 
	In the synthesis step, the excitation is first generated and then synthesized to speech through the LP synthesis filter.
	Because the variation of the excitation signal is constrained only by vocal cord movement, the training and generation processes become more efficient than those for the conventional WaveRNN vocoder, even when there are a small number of parameters.
	However, the synthesized speech is likely to be unstable, because excitation is insufficiently represented by the $\mu$-law quantization method, and the network cannot fully utilize the speech production mechanism. 
	Although the LPCNet implicitly reflects the LP synthesis process by conditioning all the signals (i.e., speech, excitation, and prediction term), it is still challenging to effectively capture their complicated interactions within a single neural network.

	To address the aforementioned limitations, we first propose an LP-mixture density network (LP-MDN), which enables the autoregressive neural vocoder to structurally represent the LP synthesis process during its training and generation processes.
	Then, we propose an improved LPCNet, i.e., iLPCNet, where the continuously defined probability distribution of speech is modeled within LP structure by incorporating the LP-MDN into the LPCNet.
	Based on the assumption that the past speech samples and the LP-related parameters are given as input of the neural vocoder, we figure out that the difference between the distributions of the speech and the excitation lies only on a constant factor represented as a \textit{prediction}\footnote{
	The term \textit{prediction} is defined as an intermediate LP operation \cite{valin2019lpcnet}, which will be discussed in detail in section~\ref{sec:lp_wavenet}.
	}. 
	If the distribution of speech is modeled by mixture of Gaussian (MoG) parameters, i.e., gain, mean, and scale components \cite{Bishop94mixturedensity, DBLP:journals/corr/abs-1711-10433}, then the distribution of the excitation can easily be converted to that of the speech by \emph{shifting} the mean component only. 
	In the proposed model, the MoG parameters of the excitation are first estimated by the WaveRNN model while the \textit{prediction} is computed from the past speech samples and the LP coefficients.
	By adding the \textit{prediction} to the mean component of the excitation, the mean component of the speech is obtained.
	Finally, the time-domain speech signal is generated by using a sampling method from its MoG distribution \cite{Bishop94mixturedensity}.

	Our contributions are summarized as follows:
	(1) We propose the LP-MDN-based iLPCNet vocoder, which enables to reliably capture the interactions between the LP filter and the excitation component.
	Because the continuous mixture output in the proposed LP-MDN does not rely on the $\mu$-law quantization process of the conventional LPCNet, the quality of the generated speech is further improved (from 4.00 to 4.41 MOS) while maintaining the model complexity of the conventional one.
	\eunwooedit{(2) We propose effective training and generation methods for improving the modeling accuracy of the iLPCNet such as a \textit{short-time Fourer transform (STFT)-based power loss function} and a \textit{distribution sharpening} that can substitute the submodules designed for the discretized waveform model of the conventional LPCNet, including pre-emphasis and waveform embedding.}

    \vspace{-2mm}
	\section{Relationship to prior work}
    \vspace{-2mm}
	There have been several attempt to incorporate the LP filter into autoregressive neural vocoding systems.
	For instance, GlotNet and ExcitNet used the WaveNet structure to generate the glottal excitation \cite{juvela2019glotnet, song2019excitnet}.
	In case of the LPCNet, it employed the lightweight WaveRNN model for fast generation of the excitation.

	Although our work is similar to those methods, its main contribution is clearly different: Our model merges the entire LP synthesis framework into both the training and generation processes.
	Note that the conventional methods require an external LP filter to synthesize the final speech waveform.
	In contrast, the proposed iLPCNet is able to directly generate a speech waveform by jointly training the interaction between the excitation and the LP synthesis filter. 
	Thus, the quality of the synthesized speech can be improved further while maintaining the LPCNet's fast generation speed.
	
	\section{LPCNet vocoder}
	\label{sec:lp_wavenet}
    The LPCNet vocoder exploits the LP-based adaptive predictor to decouple the spectral formant structure from the input speech signal. 
    Then, the WaveRNN-based generation model is used to train a distribution of the LP residual (i.e., excitation) symbols that are discretized using $\mu$-law companding.

    In the speech synthesis step, the acoustic parameters are used as input conditional features for the WaveRNN, and the model generates the corresponding time sequence of the excitation signal. 
    Finally, the speech signal is reconstructed by passing the generated excitation signal through the LP synthesis filter as follows:
	\begin{gather}
    	\label{eq:lp_exc}
    	x_{n} = e_{n} + p_n, \\ 
    	\label{eq:lp_aprox}
    	p_n = \sum_{i=1}^{M}\alpha_{i}x_{n-i}, 
	\end{gather}
	where ${x}_n$, $e_n$, and ${p}_n$ denote the $n^{th}$ sample of speech, the excitation, and the intermediate \textit{prediction} term, respectively, and $\alpha_i$ denotes the $i^{th}$ LP coefficient with the order $M$.
    In this way, the quality of the synthesized speech signal is further improved because the spectral component is well represented by the LP filter and its residual component is efficiently generated using the WaveRNN framework.

	\section{Proposed method}
	To further enhance the quality of the LPCNet with the advantages of the LP-structure, we propose an iLPCNet vocoder, in which the continuously distributed excitation signal is jointly modeled with the vocal tract filter component.
	First, we introduce the LP-MDN framework, which enables the autoregressive neural vocoder to structurally represent the LP synthesis process.
	Then, we propose the iLPCNet vocoder by applying the LP-MDN framework into the conventional LPCNet vocoder.

	\subsection{Linear prediction-structured mixture density network}
	\label{sec:lpcnetpp}
	Before describing the LP-MDN, the mathematical relationship between the probability distributions of the excitation and the speech must be clarified.
	Because both LP coefficients, $a_i$, and the previously generated sample, $x_{n-i}$, are already given components at the moment of $n^{th}$ sample generation, the prediction term, $p_n$, also can be treated as a given component.
	Thus, the difference between the probability distributions of $x_n$ and $e_n$ lies on a constant value of $p_n$, when the relationship between $x_n$ and $e_n$ shown in Eq.~\eqref{eq:lp_exc} is considered.
	
	If we define the distribution of the speech and excitation signals as a form of second-order random variable such as MoG, then the relationship between the mixture parameters of the speech and the excitation distributions can be represented as follows:
	\begin{equation}
	    \label{eq:mog}
    	p(x_n | \mathbf{x}_{<n}, \mathbf{h}) = \sum_{i=1}^{N} w_{n,i} \cdot \frac{1}{\sqrt{2\pi}s_{n,i}} \text{exp} \left[ -\frac{(x_n-\mu_{n,i})^2}{2s_{n,i}^2} \right],
	\end{equation} \vspace{-4mm}
	\begin{gather}
    	\begin{aligned}
        	\label{eq:relation3}
        	w_{n,i}^x =& w_{n,i}^e, \\	
        	\mu_{n,i}^x =& \mu_{n,i}^e   + p_n, \\
        	s_{n,i}^x =& s_{n,i}^e, 
    	\end{aligned}	
	\end{gather}
	where $x_n$, $\mathbf{x}_{<n}$, $\mathbf{h}$, and $N$ denote the current speech sample, previous speech samples, conditional acoustic features, and the number of mixtures\footnote{
	Note that if $N$ is equal to one, then the model is defined as a single Gaussian network.
	}, respectively; $[w_{n,i}$, $\mu_{n,i}$, $s_{n,i}]$ denotes the $i^{th}$ mixture parameters composed of the gain, mean, and scale components, respectively; superscripts $e$ and $x$ denote the excitation and the speech, respectively.

	Based on this observation, we propose the LP-MDN.
	First, the mixture parameters of the excitation signal are predicted using the autoregressive neural vocoder, and the prediction term, $p_n$, is computed by following Eq.~\eqref{eq:lp_aprox}.
	Then, the mixture parameters of the speech signal can be obtained as follows:
	\begin{gather}
    	\boldsymbol{w}_n = \text{softmax}(\mathbf{z}_n^{w}), \nonumber \\ 
    	\label{eq:lp_mdn:mixture2}
    	\boldsymbol{\mu}_n=\mathbf{z}^{\mu}_n + p_{n}, \\
    	\boldsymbol{s}_n = \text{exp}(\mathbf{z}_n^{s}), \nonumber
	\end{gather}
	where $[\mathbf{z}_n^w, \mathbf{z}_n^\mu, \mathbf{z}_n^s]$ denotes the output vectors of neural vocoder connected to the gain, mean, and scale nodes of the MoG distribution, respectively.
	Note that the complicated LP spectral modeling is now embedded in the mean parameters.
	In other words, the model itself presents a closed-loop solution of the spectral modeling by structurally representing the LP synthesis process.
	Thereby, the LP-MDN framework enables the neural vocoder to only focus on the excitation signal, whereas its target distribution is the entire speech signal.

	To train the network, the likelihood of MoG is first computed by following Eq.~\eqref{eq:mog}.
	Then, the weights are optimized to minimize the negative log-likelihood (NLL) loss, \ryuichiedit{$\mathcal{L}_{\mathrm{nll}}=-\sum_n \log p(x_n|\mathbf{x}_{<n}, \mathbf{h})$}.
	Because the summation of constant term guarantees linearity, the weights of the neural network can be successfully trained using a standard back-propagation process.

	\subsection{Improved LPCNet vocoder}
\begin{figure}[t!]
		\centering
		\includegraphics[width=0.9\linewidth]{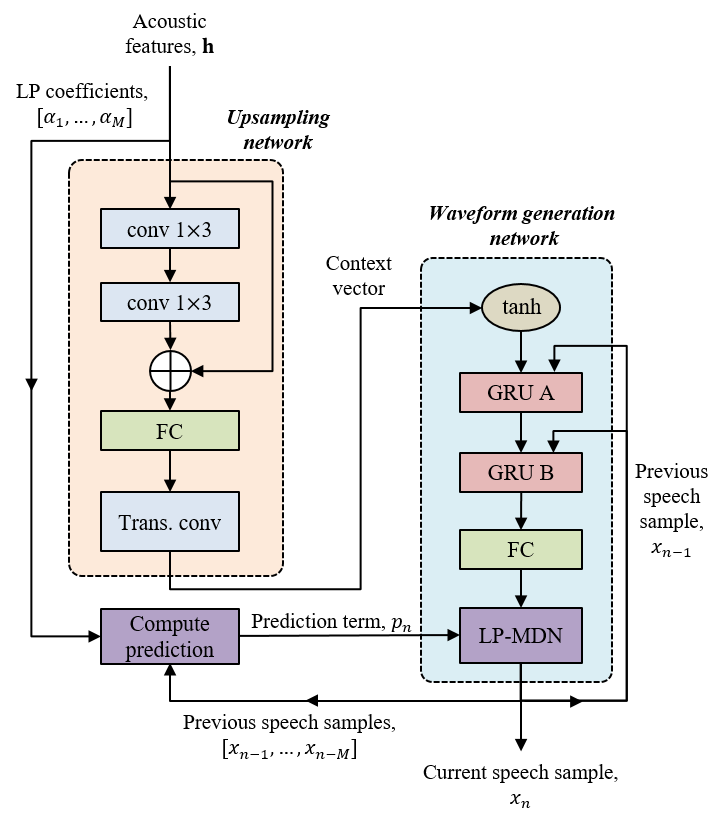}
		\caption{
			Block diagram of the proposed iLPCNet vocoder.
		}
		\label{fig:lpcnetpp}
		\vspace{-4mm}
	\end{figure}

	We propose the iLPCNet by incorporating the LP-MDN into the conventional LPCNet vocoder as illustrated in Fig.~\ref{fig:lpcnetpp}.
	Similar to the LPCNet vocoder, the iLPCNet vocoder consists of two subnetworks: the \textit{upsampling network}, which matches the time resolution of the input acoustic features to the sampling rate of the speech signal; and the \textit{waveform generation network}, which autoregressively generates the waveform samples.
	
	In the upsampling network, the local context of the acoustic features is first extracted using two 1$\times$3 convolution layers, which constructs a context vector from the current, past two, and future two frames.
	Then, the residual connection with the input acoustic features is applied to make the context vector more dominant over the current frame information.
	The fully connected (FC) layer maps the dimension of the context vector to the input dimension of the waveform generation network.
	Finally, the output of the FC layer is passed through the transposed convolution layer to upsample its time resolution.

	In the waveform generation network, the hyperbolic tangent activation is first applied to the context vectors to match their dynamic range to the waveform domain, which is bounded in $[-1, 1]$.
	Then, the concatenated vector between the context vector and the previous waveform sample goes through the two gated recurrent unit (GRU) layers and is followed by the FC layer to generate output vector, $[\mathbf{z}_n^{w}, \mathbf{z}_n^{\mu}, \mathbf{z}_n^{s}]$.
	Finally, the mixture parameters of the speech distribution are computed using the LP-MDN model, which is detailed in Eq.~\eqref{eq:lp_mdn:mixture2}.

	\subsection{Effective training and generation methods}
	As the proposed iLPCNet employs the speech distribution based on the continuously defined probability distribution, the tuning methods in the conventional $\mu$-law-based LPCNet cannot be directly applied to the iLPCNet.
	Instead, in this section, we introduce several techniques to improve the modeling accuracy of the iLPCNet.
	
	\subsubsection{STFT-based power loss}
	In addition to NLL minimization criterion, it is well known that incorporating additional auxiliary losses that are correlated with the perceptual audio quality is advantageous to improve the training efficiency \cite{ping2018clarinet}.
	In this paper, we adopt the STFT-based power loss as follows:
	\begin{equation}
	    \mathcal{L}_{\mathrm{pl}} = ||\text{STFT}(\mathbf{x}) - \text{STFT}(\hat{\mathbf{x}})||_2 ^ 2,
	\end{equation}
	where $\hat{\mathbf{x}}$ and $||\cdot||_2$ denote the generated speech sample and the L2-norm, respectively.
	Then, the weights are optimized to minimize the combined NLL and power losses as follows:
	\begin{equation}
	    \mathcal{L} = \mathcal{L}_{\mathrm{nll}} + \lambda \mathcal{L}_{\mathrm{pl}},
	\end{equation}
	where $\lambda$ is the hyper-parameter that controls the balance between $\mathcal{L}_{\mathrm{nll}}$ and $\mathcal{L}_{\mathrm{pl}}$.
	As the STFT loss function is able to effectively capture the time-frequency distribution of the realistic speech waveform, especially for the harmonic components \cite{Yamamoto2019, Yamamoto2020}, the entire training process becomes more efficient.
	\vspace{-2mm}
	
	\subsubsection{Conditional distribution sharpening for MDN model}
    Similar to the \eunwooedit{conventional} LPCNet vocoder, the \minjaeedit{noisiness of generated speech caused by the random sampling process can be controlled by adjusting the sharpness of the generated distribution.}
    In detail, the distribution is sharpened by directly reducing the generated scale parameters \eunwooedit{shown in Eq.~\eqref{eq:lp_mdn:mixture2}}.
    Because the buzziness and the hiss of synthetic speech are sensitive to the sharpness of the distribution, the scale parameters must be adjusted carefully.
    After several trials, we conclude that the best performance was presented by reducing the scale by a factor of 0.7 \ryuichiedit{in} the voiced region only.
    
    \vspace{-2mm}
    \subsubsection{Training noise injection}
    Because the \eunwooedit{conventional} LPCNet vocoder injects the training noise in the $\mu$-law domain, it is inevitable that a computational bottleneck is caused by the $\mu$-law-to-linear conversion in every training step.
    Because the proposed iLPCNet operates on the continuous linear domain, we simply add the Gaussian noise with a standard deviation of $4/2^{16}$, \eunwooedit{which corresponds to} 2-bit error at 16-bit linear quantization standard, to the conditional past speech samples in every \ryuichiedit{training} iteration.
    \vspace{-1mm}
	\section{Experiments}
	\label{sec:experiment}
	\vspace{-2mm}
	\subsection{Experimental setup}
	\subsubsection{Database}
	The experiments used a phonetically and prosodically balanced speech corpus recorded by a Korean female professional speaker.
	The speech signals were sampled at 24 kHz with 16 bit quantization.
	In total, 4,976 utterances \eunwooedit{(9.9 hours)} were used for training, 280 utterances were used for validation, and another 140 utterances were used for testing.
	
	The acoustic features were extracted using the improved time-frequency trajectory excitation vocoder at analysis intervals of every 5 ms \cite{song2017effective}, which included 40-dimensional line spectral frequencies, fundamental frequency, energy, voicing flag, 32-dimensional slowly evolving waveform, and 4-dimensional rapidly evolving waveform, all of which composed a total 79-dimensional feature vector.

	\subsubsection{Neural vocoders}
	We used the conventional WaveNet and LPCNet vocoders as baseline systems.
	In the WaveNet vocoder, the dilations were set to [$2^0, 2^1, ...,  2^9$] and repeated three times, resulting in 30 layers of residual blocks covering 3,071 samples of receptive field.
	In the residual blocks and the post-processing module, 128 channels of convolution layers were used.
	The target speech signal was quantized from 0 to 255 using 8-bit $\mu$-law compression.

	In the LPCNet vocoder, the \minjaeedit{specification was almost \eunwooedit{the} same} as its original version \cite{valin2019lpcnet}.
	In the frame-level network, the dimension of the two FC layers was set to 64.
	In the sample-level network, 256- and 16-dimensional GRU layers were used.
	In addition, 256-dimensional waveform embedding matrices and dual FC layer were used.
	The softmax distribution in the voiced region was sharpened by multiplying factor of two to the logits of softmax distribution \cite{zeyu2018FFTNet}.

	In the proposed iLPCNet vocoder, the 256-dimensional FC layer was used in the upsampling network.
	To compose the transposed convolution, the kernel size and stride interval were set to 120 (5 ms).
	In the waveform generation network, the dimensions of the first and second GRUs were set to 256 and 16, respectively.
	For the simplicity, the distribution of the speech signal was defined as single Gaussian; therefore, the output dimension of the last FC layer was two.
	The weight value for power loss, $\lambda$, was set to 10.
	The weight normalization technique was applied to each convolution and FC layer to stabilize the training process \cite{salimans2016weight}.
	Note that the LPCNet and the iLPCNet had the same GRU dimensions and that their generation complexities were nearly the same.
	
	The weights of all the neural vocoders were first initialized using the \textit{Xavier} initializer \cite{xavier2010init}, and then trained using an \textit{Adam} optimizer \cite{diederik2014adam}.
	The Noam scheme-based learning rate scheduling method having a base learning rate of $10^{-3}$ and a warm-up step of 4,000 iterations was applied \cite{vaswani2017attention}.
	The mini-batch size was 10,000 samples with 8 GPUs, resulting in 80,000 samples per mini-batch.
	The networks were trained in 50 epochs, which corresponds to about 530,000 iterations.
    
    \vspace{-2mm}
	\subsubsection{Acoustic model for text-to-speech}
    To evaluate the vocoding performance in the text-to-speech (TTS) framework, a Tacotron 2 model was used as an acoustic model \cite{Shen2018NaturalTS}.
    In the training process, the input text sequence was first converted into 512-dimensional character embeddings by using a front-end character embedding module\footnote{ 
    A sub-character architecture was used to design the front-end character module to process inputted Korean text sequences \cite{stratos2017sub}.
    }, and then fed into a sequence-to-sequence transcript encoder.
    The transcript encoder had three convolution layers with 10$\times$1 kernel and 512 channels, and the final layer was followed by a bi-directional long short-term memory (LSTM) network that had 512 memory blocks.

	\begin{table}[t!]
		\caption{MOS test result with a 95\% confidence interval for various speech synthesis systems: Acoustic features extracted from the recorded speech (A/S) and generated from the acoustic model (TTS) were used to compose the input auxiliary features.}
		\setlength\tabcolsep{4.2pt} 
		\label{table:subjective_mos}
		\renewcommand{\arraystretch}{1.0}
		\centering
	    \small
		\begin{tabular}{c||c|c}
			\Xhline{2\arrayrulewidth}
			System & A/S & TTS  \\ \hline\hline
			WaveNet & 3.29$\pm$0.19 & 3.33$\pm$0.13 \\
			LPCNet & 4.05$\pm$0.21 &  4.00$\pm$0.11  \\
			iLPCNet (ours) & 4.34$\pm$0.14 & 4.41$\pm$0.18 \\ \hline
			Recorded & \multicolumn{2}{c}{4.71$\pm$0.09} \\ \Xhline{2\arrayrulewidth}
		\end{tabular}
		\vspace{-4mm}

	\end{table}

    To align the transcript embedding with the target acoustic features, a context vector containing both location and content information was extracted by using a location-sensitive attention network \cite{chorowski2015attention}, which had convolution layers comprising 64$\times$1 kernel followed by a fully connected projection layer having 128 units.

	To decode the acoustic features, the previously generated features were fed into two FC layers having 256 units to extract the bottleneck features.
    Then, these bottleneck features and the context features from the attention network were passed through uni-directional LSTM layers that had 1,024 memory blocks and two projection layers, which generated a stop token and the acoustic features.
    Finally, five convolution layers having 5$\times$1 kernel and 512 channels were used as a post-processing network to add the residual component of the generated acoustic features.
    
    To train the network, the weights were optimized to minimize the criteria of the mean square error between the acoustic features extracted from the recorded speech and estimated from by the predicted acoustic features.
    The learning rate was scheduled for decay from $10^{-3}$ to $10^{-4}$ via a decay rate of 0.33 for each 100,000 steps.

	\begin{figure}[t!]
		\hspace{-6mm}
		\includegraphics[width=1.10\linewidth]{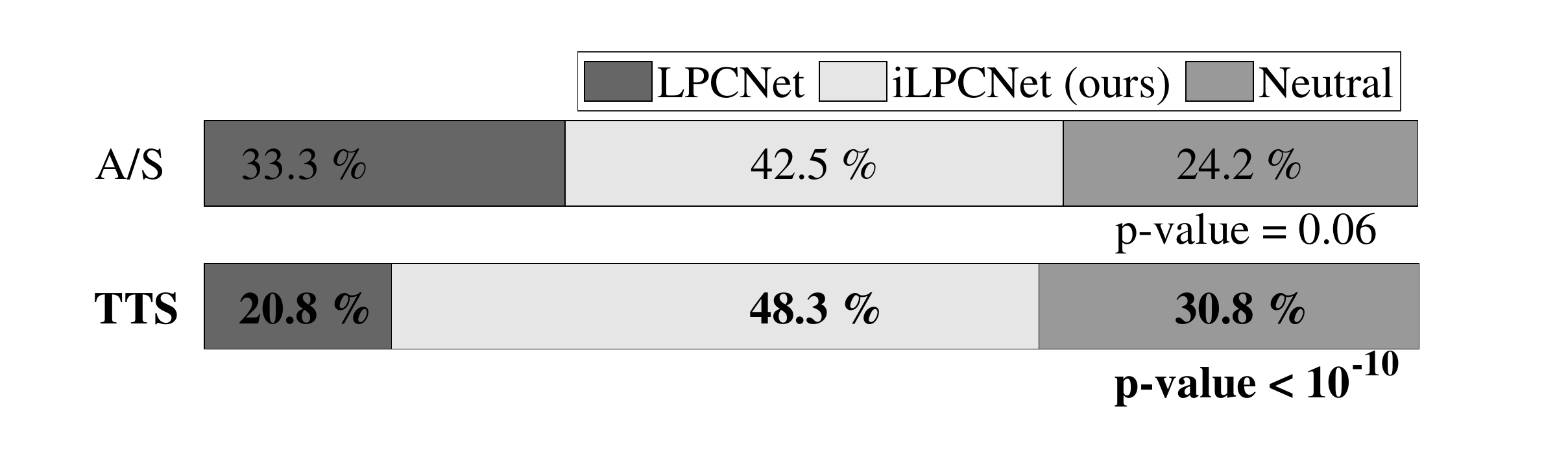}
	    \vspace{-9mm}
		\caption{
			Subjective preference test results (\%) between conventional LPCNet and proposed iLPCNet.
			The systems that achieved significantly better preference at the $p < 0.01$ level are in bold typeface.
		}
	    \vspace{-2mm}

		\label{fig:pref_test}
	\end{figure}
	
    \subsection{Evaluations}
    \vspace{-2mm}
	To evaluate the perceptual quality of the proposed system, the mean opinion score (MOS) listening test and the A-B preference test were performed\footnote{Generated audio samples are available at the following URL:	\\ \url{https://min-jae.github.io/icassp2020/}}.
	In both tests, 15 native Korean listeners were asked to rate the performance of 15 randomly selected synthesized utterances from the test set.

	Table~\ref{table:subjective_mos} summarizes the MOS test results, whose trends can be analyzed as follows:
	(1) In both the analysis and synthesis (A/S) and TTS cases, the LPCNet-based neural vocoders performed significantly better than the WaveNet-based one.
	This confirms that decoupling the spectral component of the speech signal benefits the modeling accuracy of the remaining signal.
	(2) Among the vocoders with LP filters, the proposed iLPCNet showed a higher perceptual quality than the conventional LPCNet, implying that the LP-MDN model helped to reconstruct more natural speech signals.
	(3) Consequently, the proposed iLPCNet with the Tacotron 2 acoustic model achieved a 4.41 MOS.

	The preference test results shown in Fig.~\ref{fig:pref_test} also verify that the listeners preferred the proposed iLPCNet over the conventional LPCNet.
	The results were significant in the TTS framework, implying that the proposed iLPCNet can robustly synthesize the speech waveform even though the acoustic parameters contain prediction errors.

    \vspace{-1mm}
    
	\section{Conclusion}
	\label{sec:conclusion}
	\vspace{-2mm}
	This paper proposed the iLPCNet vocoder by applying LP-MDN to the conventional LPCNet model.
	By utilizing the causality of the LPCNet and the linearity of the LP filter, we structurally merged the LP synthesis process into the LPCNet framework.
	The proposed iLPCNet enabled generation of the high-resolution speech signals via the LP-MDN model, and the quality of the generated speech was better than that of the conventional LPCNet.
	Furthermore, the pipelines of the LPCNet was made simpler and more compact by the removal of the additional modules designed for signals quantized by $\mu$-law.
	The experimental results verified that the proposed system significantly outperformed the conventional neural vocoding systems.
    
	\section{Acknowledgements}
	\vspace{-2mm}
    The work was supported by Clova Voice, NAVER Corp., Seongnam, Korea.

	\vfill
	\pagebreak
    \bibliographystyle{IEEEtran}
    \bibliography{mybib_mj}
\end{document}